\pdfoutput=1
\documentclass[aps,twocolumn,floatfix,prb]{revtex4}
\usepackage[final]{graphicx}
\DeclareGraphicsExtensions{.pdf}
\usepackage{amsmath,amssymb,bbold,bm}
\usepackage{float}

\newcommand{\bE}{{\bf E}}
\newcommand{\bB}{{\bf B}}
\newcommand{\bA}{{\bf A}}
\newcommand{\bx}{{\bf x}}

\newcommand{\cC}{{\cal C}}
\newcommand{\cT}{{\cal T}}
\newcommand{\cP}{{\cal P}}

\begin{document}

\title{Quantization and $2\pi$ Periodicity of the Axion Action in Topological Insulators}

\author{M.M. Vazifeh}
\author{M. Franz}
\affiliation{Department of Physics and Astronomy, University of
British Columbia, Vancouver, BC, Canada V6T 1Z1}

\begin{abstract}
The Lagrangian describing the bulk electromagnetic response of a
three-dimensional strong topological insulator contains a topological
`axion' term of the form $\theta\bE\cdot\bB$. It is often stated
(without proof) that the corresponding action is quantized on periodic
space-time and therefore invariant under $\theta\to\theta +2\pi$. Here
we provide a simple, physically motivated proof of the axion action
quantization on the periodic space-time, assuming only that the vector
potential is consistent with single-valuedness of the electron
wavefunctions in the underlying insulator. \end{abstract}

\date{\today}

\maketitle

{\em Introduction ---}
Topological insulators are time-reversal $(\cT)$ invariant crystalline
solids with insulating bulk band structure and topologically protected
gapless surface states.\cite{mooreN,hasan_rev} These surface states
are robust against the effects of non-magnetic disorder and form a
theoretical basis for numerous exotic
phenomena\cite{fu3,qi1,seradjeh1,qi2} as well as proposed practical
applications.\cite{nagaosa1,garate1} Their presence has been detected
in several materials with a strong spin-orbit
interaction.\cite{sczHgTe2, hasan1,hasan2,shen1,hasan3}

An alternative characterization of a strong topological
insulator\cite{fu-kane3D,moore1,roy1} (STI) follows from its unusual
response to applied electromagnetic fields which is encoded in a bulk
`axion' term\cite{qi1,moore2} of the form
\begin{equation}\label{ax} {\cal L}_{\rm axion}=\theta \left({e^2\over
2\pi hc}\right)\bB\cdot\bE,
\end{equation}
with $\theta=\pi$. The axion term (\ref{ax}) appears in the
electromagnetic Lagrangian in addition to the standard Maxwell
term. Eq.\ (\ref{ax}) underlies the topological magnetoelectric
effect\cite{qi1,moore2} in which electric (magnetic) polarization is
induced by external magnetic (electric) field, as well as the Witten
effect\cite{witten1,marcel2} that attaches a fractional electric
charge to a magnetic monopole.

The axion term (\ref{ax}) has been introduced in the context of
high-energy physics decades before the discovery of STIs to resolve CP
non-violation problem in quantum
chromodynamics\cite{pecceiAx,weinberg0,wilczek0} (QCD). The
corresponding $\theta(\bx,t)$ field is known to particle physicists as
the axion field\cite{wilczekAx}. The action of the uniform axion field
can be viewed as a topological term for the $\theta$ vacuum in QCD
arising from nontrivial topology of such a vacuum.

For a generic value of $\theta$ the axion term breaks $\cT$ as well as
parity $\cP$. This is because under time-reversal $\bB\to-\bB$,
$\bE\to \bE$, while under spatial inversion $\bB\to\bB$, $\bE\to
-\bE$. What allows the $\cT$- and $\cP$-invariant insulators to
possess an axion term with $\theta=\pi$ is the $2\pi$-periodicity of
the axion action $S_{\rm axion}=\int dt d^3x {\cal L}_{\rm axion}$ in
parameter $\theta$. Specifically, on periodic space-time (that is used
to model an infinite bulk crystal), the integral in the axion action is
quantized,
\begin{equation}\label{ax2}
 \left({e^2\over 2\pi hc}\right) \int dt d^3x \bB\cdot\bE = N\hbar,
\end{equation}
with $N$ integer. All physical observables depend on $\exp{(iS_{\rm
    axion}/\hbar)}$ and are thus  invariant under a global
transformation $\theta\to\theta+2\pi$. Consequently, $\theta=\pi$ and
$\theta=-\pi$ are two equivalent points and describe a $\cT$- and
$\cP$-invariant system. Conversely, in a system invariant under $\cP$
or $\cT$ the value of $\theta$ is quantized to 0 or $\pi$.

The statement regarding the quantization of the expression (\ref{ax2})
on periodic space-times has been made in several influential papers
\cite{qi1,qi2,moore2,wilczekAx} but no simple physical explanation has been given of
its validity. One way to understand the quantization is using mathematical theory of 
fibre bundles\cite{nakahara} in which the axion action is an integral of a second Chern character 
associated with an abelian gauge theory. In topology Chern characters are 
forms whose integral over closed base space returns integer values. Hence, once the base space 
is compact the topological axion action is necessarily quantized. A well-known 
example in the context of condensed matter physics is the transverse conductivity of a quantum                          
Hall insulator which can be expressed as a first Chern integral over the BZ and hence it turns out to be strictly quantized.\cite{TKNN, ChernQH} 

Since the quantization of the axion term and the related $\theta$-periodicity 
underlies the essential element of the theory of topological insulators it is important 
to have a clear physical understanding of its origin. In the
rest of this paper we provide a direct and simple proof of the axion
action quantization on periodic space-time. We also consider a non-periodic case where the axion action remains quantized.                                                                                                                              
Our proof is based on the electromagnetic field decomposition into an
`externally imposed' uniform constant part which we show can have
non-zero contribution to $S_{\rm axion}$ and a part generated by
space-time periodic charge and current configurations whose
contribution vanishes. The quantization condition (\ref{ax2}) follows
from the requirement that the underlying vector potential be
consistent with the single-valuedness of the electron
wave-functions.\cite{aharonov-bohm} In our proof we assume that no
magnetic monopoles are present but in closing we comment on the situation with monopoles. As a byproduct of our proof we 
find that for abelian gauge field the invariant based on the second Chern 
character is fully determined by first Chern characters associated with 
orthogonal coordinate planes in four-dimensional space-time. In this way 
abelian case is different from the better known non-abelian case\cite{nakahara} 
where due to the presence of instantons the second Chern 
number can be non-zero even if all first Chern numbers vanish.

{\em General considerations ---}
In a covariant formulation with the speed of light $c=1$ the axion
action can be written as \cite{wilczekAx}

\begin{equation}\label{Axion1} \frac{1}{\hbar}S_{\text{axion}} =
\frac{\theta}{8\Phi_{0}^2}\int{d^{4}x\;
\varepsilon^{\mu\nu\alpha\beta} F_{\mu \nu} (x)F_{\alpha \beta}}(x)
\end{equation}
where $F_{\mu\nu} = \partial_{\mu} A_{\nu} - \partial_{\nu} A_{\mu}$
is the electromagnetic field tensor and $\Phi_{0}={h / e}$ is the
quantum of magnetic flux. In the following we consider a space-time
hypercube of side $L$ with periodic boundary conditions imposed on
$F_{\mu \nu} (x)$ in all directions.

\begin{figure}[t]
\includegraphics[width=8cm]{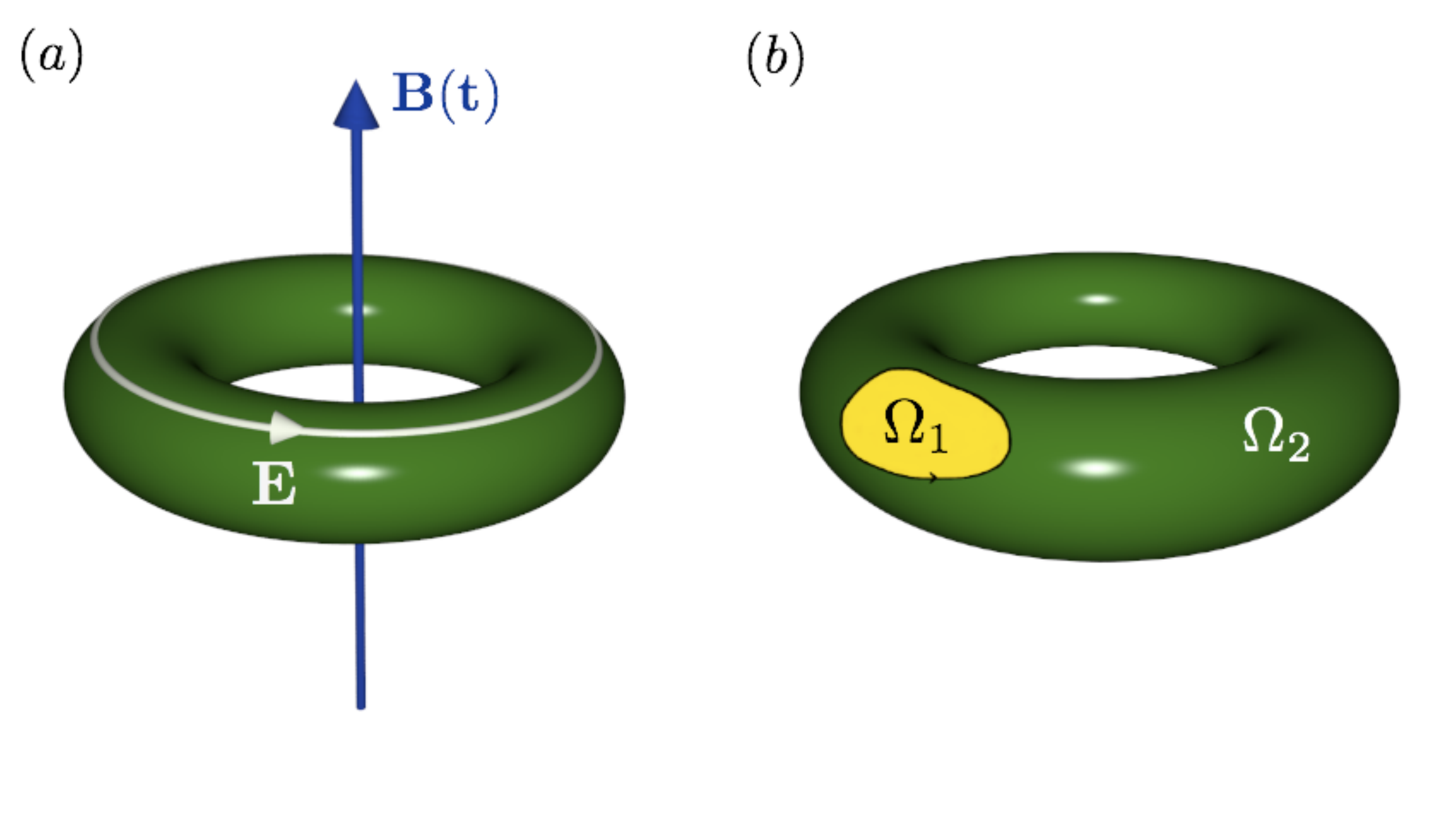}
\caption{(Color online) (a) A magnetic field increasing linearly with
time and confined to a torus hole produces a uniform and constant
electric field along the torus.  (b) A closed path on the torus can be
thought of as enclosing two areas, $\Omega_1$ and $\Omega_2$.} \label{fig1}
\end{figure}

In the absence of monopoles integration by parts gives
\begin{equation}\label{Axion2}
\frac{1}{\hbar}S_{\text{axion}}=\frac{\theta}{4\Phi_{0}^2}\int{d^{4}x\;
\varepsilon^{\mu\nu\alpha\beta}\partial_{\alpha} \left[ F_{\mu \nu}(x)
A_{\beta}(x)\right]}
\end{equation}
At first glance, from the periodicity of space and time one might
conclude that the integral in (\ref{Axion2}) vanishes for a general
electromagnetic field tensor since it can be written as a three
dimensional hyper-surface integral of $F_{\mu \nu} A_{\beta}$ which is
zero if this function is periodic in space-time coordinates. However,
a simple example of constant uniform fields $\bE\|\bB$ shows this
conclusion to be erroneous. The point is that in general the gauge
field 4-vector of a periodic electromagnetic field is not periodic in
space and time. As an example consider a lower dimensional
case of $T^2$ torus with a magnetic flux through its hole increasing
linearly with time (Fig.\ \ref{fig1}a). This induces an electric field
on the torus which is constant and therefore periodic in time and the
coordinates that parametrize the torus. However the line integral of
the gauge field over the non-contractible loop enclosing the magnetic
flux is nonzero which means that the gauge field cannot be chosen
periodic. For the field configurations of this type, containing field
lines along non-contractible loops, $S_{\text{axion}}$ will be
non-vanishing and we must consider these with special care.

A question arises here: in general for what kind of periodic
electromagnetic field configurations in 3 spatial dimensions the gauge
field cannot be chosen periodic? A gauge potential $A_\mu$ cannot be
chosen periodic if
\begin{equation}\label{A2} \oint_{\square} d \ell \cdot A\neq 0
\end{equation}
where the line integral is over any $L \times L$ square located in one
of the coordinate planes $x_\mu x_\nu$ with $\mu\neq\nu$ on the
hyper-cube. We note that the above integral (\ref{A2}) is gauge
invariant and, due to the periodicity of $F_{\mu\nu}(x)$, its value is
independent of the position of the $L \times L$ square, i.e. it is
invariant under any space-time translation. In terms of the field
tensor Eq.\ (\ref{A2}) can be written as
\begin{equation}\label{A3} \int_{\square} dx_\mu dx_\nu F_{\mu\nu}
\neq 0,
\end{equation}
with no summation over $\mu$, $\nu$.  Physically, this means that
total magnetic flux through one of the spatial faces of the hypercube
is non-zero and a similar condition for the electric field (see
below).

The above considerations motivate a decomposition of the gauge
potential into two pieces,
 \begin{equation}\label{A1}
A_{\mu}(x) = A^{0}_{\mu}(x) + \delta A_{\mu}(x),
\end{equation}
such that 
\begin{equation}\label{A4}
\oint_{\square} d \ell \cdot \delta A=0.
\end{equation}
Hence $\delta A_\mu$ can be chosen periodic, while $A^{0}_{\mu}(x)$ contains 
any non-periodic part. Similarly, we write 
\begin{equation}\label{F4}
F_{\mu\nu}(x) = F_{\mu\nu}^{0}(x) + \delta F_{\mu\nu}(x),
\end{equation}
with $F_{\mu\nu}^{0}\equiv \partial_{\mu} A^0_{\nu} - \partial_{\nu} A^0_{\mu}$.
To make this decomposition unique (up to a gauge transformation) we
furthermore impose a condition that components of $F_{\mu\nu}^{0}$ are
uniform and constant,
\begin{equation}\label{f3}
F_{\mu\nu}^{0}={1\over L^2}\int_{\square} dx_\mu dx_\nu F_{\mu\nu}.
\end{equation}
In terms of electric and magnetic field vectors our decomposition corresponds to 
\begin{equation}\label{EB}
\bE = \bE_{0}+\delta \bE,\;\;\; \bB = \bB_{0}+\delta \bB
\end{equation}
where $\bE_{0}$, $\bB_{0}$ are constant uniform fields while $\delta
\bE$, $\delta \bB$ are space-time varying fields derived from the
periodic gauge potential $\delta A_\mu(x)$. The constant fields can
only be produced by magnetic and electric fluxes through the holes in
the $T^3$ torus embedded in a four dimensional space and it is not
possible to devise non-singular charge and current sources within the
periodic three dimensional space to produce them.\cite{jackson}

Fields $\delta \bE$ and $\delta \bB$ are produced by ordinary charge
and current sources. They have the following physical properties:
(i) Magnetic fluxes associated with these magnetic fields through each
spatial face vanish,
\begin{equation}\label{Bflux} \varepsilon^{ijk}\int_{\square} d x_i
dx_j \delta B_k = 0 \;\;\;\;\;\; i,j,k=1,2,3.
\end{equation}
There is no summation on indices and the equation holds for all $x_k$
and $x_0$.
(ii) The integral over any space-time face
\begin{equation}\label{Eintegral}
\int_{\square} d x_0 d x_i \delta E_i = 0,
\end{equation}
with no summation on $i = 1,2,3$.
(i) and (ii) are properties of the fields produced by a general
space-time periodic charge and current sources.

{\em Action evaluation ---} With this preparation we can now proceed
to evaluate the axion action. It is most convenient to employ Eq.\
(\ref{Axion2}) where in view of our decomposition (\ref{A4},\ref{F4})
the expression $F_{\mu\nu}A_{\beta}$ is replaced by
\begin{equation}\label{FA} F^0_{\mu\nu}A^0_{\beta}+ 2
F^0_{\mu\nu}\delta A_{\beta} + \delta F_{\mu\nu}\delta A_{\beta}.
\end{equation}
An integration by parts has been performed on $\delta
F_{\mu\nu}A_{\beta}^0$ to obtain the factor of 2 in the middle
term. Now the second and the third term in the above expression
(\ref{FA}) are explicitly space-time {\em periodic} and therefore
their contribution to $S_{\text{axion}}$ identically vanishes. The
only contribution to the action comes from the first term which
represents the uniform constant part of the electromagnetic
fields. Thus,
\begin{equation}\label{Axion3} \frac{1}{\hbar}S_{\text{axion}} =
\frac{\theta}{\Phi_{0}^2}\int{d^{4}x\; \bE_0\cdot\bB_0}.
\end{equation}
It remains to be demonstrated that the action is quantized for these
constant and uniform fields.

Our arguments thus far have been purely classical. At the level of
classical electrodynamics, clearly, the integral in Eq.\
(\ref{Axion3}) can attain any desired value and is not quantized.  To
proceed, we must recall that in the present context the axion term
results from integrating out the electron degrees of freedom in a
topological insulator. Electron behavior is inherently quantum
mechanical. The axion action quantization then follows from the
requirement that the gauge potential $A_\mu$ that couples to the
electron wavefunctions be consistent with the quantum theory of
electrons in periodic space-time.

In the following we assume for simplicity that our fields are pointed
along the $x_3$ direction, $\bE_0=E_3\hat{x}_3$ and
$\bB_0=B_3\hat{x}_3$. Other components can be treated in an identical
fashion. For this configuration we may decompose our space-time torus
$T^4$ into a direct product $T^2_{12}\times T^2_{03}$ and write
\begin{equation}\label{Axion4} \frac{1}{\hbar}S_{\text{axion}} =
\frac{\theta}{\Phi_{0}^2}\int_\square{dx_1 dx_2 B_{3}} \int_{\square}{
dx_0dx_3 E_{3}}.
\end{equation}
It remains to show that each of these integrals 
is an integer multiple of magnetic flux quantum $\Phi_0$. The first integral represents the 
total magnetic flux through the $x_1x_2$ face of the hypercube. The quantization 
of this term follows from the standard arguments for the electron motion in applied 
magnetic field, which we now briefly review for completeness. 

Imagine an arbitrary
closed path $\cC$ on the $T^2_{12}$ torus. As illustrated in Fig.\ \ref{fig1}b it encloses area denoted as $\Omega_1$. Alternately, it can be viewed as enclosing its complement on $T^2_{12}$ denoted as $\Omega_2$. Using Stoke's theorem we may write 
\begin{eqnarray}\label{Axion77a} 
\int_{\Omega_1}\bB\cdot d{\bf S}&=&\oint_\cC\bA\cdot d{\bf l} \\
\int_{\Omega_2}\bB\cdot d{\bf S}&=&-\oint_\cC\bA'\cdot d{\bf l}\label{Axion77b} 
\end{eqnarray}
where the prime on the vector potential signifies the subtle but important fact that the equality is required to hold only up to a gauge transformation 
$A_{\mu} \rightarrow A_\mu'=A_{\mu} - \partial_{\mu} f$, with $f(x)$ a scalar function. Now the line integral of $\bA$ along a closed path is normally thought of as a gauge invariant quantity in which case adding Eqs.\ (\ref{Axion77a}) and (\ref{Axion77b}) immediately implies $\int_{\Omega_1+\Omega_2}\bB\cdot d{\bf S}=0$. This suggests that $S_{\rm axion}/\hbar$ is indeed quantized but the only value allowed is 0. However, there exists a class of `large' gauge transformations $f(x)$ which change the value of the line integral but leave the wavefunction single valued. The latter transforms as $\Psi(x) \rightarrow \Psi'(x)=e^{i e f(x)}\Psi(x)$ and the relevant $f(x)$ contains a vortex (a Dirac string) at some point of the $T^2_{12}$ torus, i.e. $e^{i e f(x)}\sim e^{i e n\varphi}$ where $\varphi$ is an angle in $x_1x_2$ plane measured from the vortex center and $n$ is an integer. Since $\oint_\cC\nabla f \cdot d{\bf l}=2\pi n$ the inclusion of large gauge transformations of this type can be seen from Eqs.\ (\ref{Axion77a}) and (\ref{Axion77b}) to allow for non-zero quantized values
$\int_\square{dx_1 dx_2 B_{3}}=n\Phi_0$.

%

One can advance the same argument to establish the quantization of the second 
surface integral in Eq.\ (\ref{Axion4}). Consider a closed path, this time on $T^2_{03}$, enclosing $\Omega_1$ and $\Omega_2$ regions. It is straightforward to check that all steps proceed exactly as before. The large gauge transformations now involve space-time vortices in $f(x)$ (i.e.\ vortices in the $x_0x_3$ plane) and lead to analogous result $\int_\square{dx_0 dx_3 E_{3}}=m\Phi_0$ with $m$ integer.

Combining the above results we find
\begin{equation}\label{Axion5} \frac{1}{\hbar}S_{\text{axion}} =
N\theta,
\end{equation}
with $N=nm$.  Eq.\ (\ref{Axion5}) shows that the axion action for
electromagnetic field is quantized on periodic space-time and,
consequently, the amplitude $\exp(iS_{\text{axion}}/\hbar)$ is
invariant under the shift of the axion angle $\theta$ by any integer
multiple of $2\pi$.

A more abstract and rigorous  way to think of the quantization of these quantities is using fibre bundle 
mathematics mentioned briefly in the beginning. The fibre bundles\cite{nakahara} can be classified 
in terms of the so called Chern characters. These are forms whose integral over closed base space of the 
bundle always return an integer number. One consequence is that the integral of a first Chern character of 
the Abelian $U(1)$ gauge theory, $F_{\mu\nu}/\Phi_0$ on the closed base space $T^2_{\mu\nu}$ must 
always be an integer. On the other hand the axion action is a second Chern character integral evaluated 
for this abelian gauge theory. We showed that this can be written as a product of two first Chern character 
integrals whose quantization, as we discussed, has a clear physical interpretation.

{\em Non-periodic systems ---} Assuming periodic boundary conditions
in all directions is the simplest way to avoid edges and to
concentrate on the bulk response.  However, in real experimental setup
one must deal with a situation where the the fields are present in a
finite portion of space and over a finite time duration. The question
arises whether the axion action remains quantized under these
non-periodic conditions. The answer is ``yes'' provided that one
additional condition on the gauge potential is
satisfied. Specifically, it is possible to show that Eq.\ (\ref{ax2})
remains valid if (i) the fields $\bB$ and $\bE$ vanish outside a
space-time volume ${\cal V}$ and (ii) the underlying gauge field
$A_\mu$ is such that its presence cannot be detected by any
Aharonov-Bohm type experiment performed using charge $e$ particles
outside ${\cal V}$. For the magnetic field this implies, for example,
that the total flux enclosed by any closed trajectory is $n\Phi_0$
with $n$ integer. In that case the Aharonov-Bohm phase acquired by
charge $e$ particle is $2\pi n$ and thus indistinguishable from 0.

{\em Inclusion of monopoles ---} In passing from Eq.\ (\ref{Axion1})
to (\ref{Axion2}) we assumed that a term
$A_\beta\varepsilon^{\mu\nu\alpha\beta}\partial_\alpha F_{\mu\nu}$
that appears in the integration by parts vanishes on the account of
partial derivatives commuting and $\varepsilon^{\mu\nu\alpha\beta}$
being antisymmetric. This assumption fails in the presence of magnetic
monopoles. Consider {\em e.g.}\ the $\beta=0$ component of the above
expression which equals $2A_0\nabla\cdot\bB$.  In the presence of the
non-vanishing monopole density $\nabla\cdot\bB\neq 0$ such term will
give non-zero contribution to $S_{\text{axion}}$ whenever $A_0$ is
non-zero. Similarly, $\beta=1,2,3$ terms correspond to monopole
currents and may be non-vanishing as well. Our proof of
$\theta$-periodicity given above must be modified in the presence of
monopoles.

The simplest modification applies to the special case when there are
no electrical charges or currents in the system and fields are sourced
purely by magnetic charges and currents. In this case one can perform
a duality transformation\cite{jackson} which interchanges $\bE$ and
$\bB$ and the proof proceeds exactly as before in terms of dual field
variables.
In the most general case when there are both electric and magnetic
charges/currents present a form of $\theta$-periodicity still holds
but its statement and proof now involve several subtle
points.\cite{vazifeh1} Specifically one must take special care when
dealing with non-single valued vector potential and one must also take
into account the Witten effect.\cite{witten1}

{\em Closing thoughts ---} We have presented a simple and intuitive
proof of the quantization of the topological axion action on periodic
space-time.  Our considerations show that the theory is invariant
under a global $\theta \rightarrow \theta +2\pi$ transformation
consistent with the $\mathbb{Z}_2$ character of the fundamental
`strong' invariant describing the physics of time-reversal invariant
band insulators.
\\
\\
\emph{Acknowledgment} ---  The authors benefited greatly from the discussions and correspondence with I. Affleck, 
H. Karimi, J.E. Moore, G. Rosenberg, G.E. Volovik, X.-L. Qi and G. Semenoff. Support for this work came from NSERC and CIfAR.

\end{document}